\documentclass[twocolumn,showpacs,aps,amsmath,amssymb,floatfix,superscriptaddress]{revtex4}
\usepackage{graphicx}
\begin{document}
\title{Weak Disorder in Fibonacci Sequences}
\author{E.~Ben-Naim}
\email{ebn@lanl.gov} 
\affiliation{Theoretical Division and Center for Nonlinear Studies,
Los Alamos National Laboratory, Los Alamos, New Mexico 87545 USA}
\author{P.~L.~Krapivsky} 
\email{paulk@bu.edu}
\affiliation{Department of Physics and Center for Molecular
Cybernetics, Boston University, Boston, Massachusetts, 02215 USA}
\begin{abstract}
We study how weak disorder affects the growth of the Fibonacci
series. We introduce a family of stochastic sequences that grow by
the normal Fibonacci recursion with probability $1-\epsilon$, but
follow a different recursion rule with a small probability
$\epsilon$. We focus on the weak disorder limit and obtain the
Lyapunov exponent, that characterizes the typical growth of the
sequence elements, using perturbation theory. The limiting
distribution for the ratio of consecutive sequence elements is
obtained as well. A number of variations to the basic Fibonacci
recursion including shift, doubling, and copying are considered.
\end{abstract}
\maketitle

The Fibonacci integer sequence $\{1,1,2,3,5,8,\ldots\}$ has been
studied extensively in number theory, applied mathematics, physics,
computer science, and biology \cite{cg,http}. Fibonacci numbers are
ubiquitous in nature: they govern branching in trees, spiral
patterns in shells, and the arrangement of seeds in sunflowers
\cite{wh,rvj,dc,ns}.

The Fibonacci sequence, defined via the recursion relation
\begin{equation}
\label{normal} F_{n+1}=F_{n}+F_{n-1}
\end{equation}
with $F_0=0$ and $F_1=1$, is deterministic. However, many patterns
in nature are not perfect. For example, spiral patterns in
sunflowers, where Fibonacci numbers as high as 144 are observed, may
very well be disordered. An empirical study of sunflowers observes
the normal sequence $\{1,1,2,3,5,\ldots\}$ with a frequency of 95\%,
but altered sequences such as $\{2,3,5,7,\ldots\}$ and
$\{1,3,4,\ldots\}$ are also observed with a small frequency
\cite{rvj}.

Motivated by this empirical observation, we study disorder in
Fibonacci sequences. Specifically, we introduce the following {\it
stochastic} sequence. We assume that the normal Fibonacci rule
(\ref{normal}) is followed most of the time, but that with a small
probability, $\epsilon\ll 1$, an ``erroneous'' recursion
$x_{n+1}=x_n+x_{n-2}$, involving an index shift, is followed. The
stochastic recursion rule is therefore
\begin{equation}
\label{def}
x_{n+1}=
\begin{cases}
   x_{n}+x_{n-1}   & {\rm prob} \quad 1-\epsilon;\\
   x_{n}+x_{n-2}   & {\rm prob} \quad \epsilon.
\end{cases}
\end{equation}
The initial elements are always $x_0=0$, and $x_1=x_2=1$.

Let us first recall a few useful facts on the Fibonacci sequence,
corresponding to the limiting case $\epsilon=0$. The series elements
are given by
\begin{equation}
\label{fn}
F_n=\frac{\lambda^n-(-\lambda)^{-n}}{\lambda+\lambda^{-1}}
\end{equation}
with the golden ratio $\lambda=\frac{1+\sqrt{5}}{2}$. The series
elements grow exponentially with $n$, $F_n\sim \lambda^n$.
Substituting this form into the recursion (\ref{normal}), the golden
ratio satisfies \hbox{$\lambda^2=\lambda+1$}, and this allows to
express polynomials of arbitrary degree in $\lambda$ as linear
functions of $\lambda$. Moreover, the ratio between two successive
series elements, $r_n=x_n/x_{n-1}$, approaches the golden ratio
$r_n\to \lambda$, as $n\to\infty$.

Our goal is to elucidate the typical growth of the sequence elements
\begin{equation}
\label{typical}
x_n\sim e^{\beta n}
\end{equation}
with $\beta\equiv\beta(\epsilon)$ the Lyapunov exponent. For
example, for the random Fibonacci series $x_n=x_{n-1}\pm x_{n-2}$
where addition and subtraction are chosen with equal probabilities,
the Lyapunov exponent is $\beta\approx 0.123975$
\cite{hf,et,dv,sk,mm}. In our case, however, the recursion rules
(\ref{def}) represent a gentle departure from the original Fibonacci
rule (\ref{normal}) and thus, we expect a small change in the
Lyapunov exponent. We focus on the weak disorder limit, $\epsilon\to
0$, and use perturbation theory to show that the Lyapunov exponent
varies linearly with the disorder strength
\begin{equation}
\label{pert}
\beta(\epsilon)=\beta_0+\beta_1\epsilon+\cdots
\end{equation}
with $\beta_0=\ln \lambda$.

In general, the average behavior $\langle x_n\rangle$ can be
obtained analytically. From the sequence definition (\ref{def}),
the average satisfies the recursion relation
\begin{equation}
\label{av}
\langle x_{n+1}\rangle= \langle
x_n\rangle+(1-\epsilon)\langle x_{n-1}\rangle+\epsilon\langle
x_{n-2}\rangle
\end{equation}
with $\langle x_0\rangle=0$, and $\langle x_1\rangle=\langle
x_2\rangle=1$. This linear relation implies the exponential growth
$\langle x_n\rangle\sim \mu^n$ with the growth factor $\mu$ being
the largest root of the third order polynomial
\begin{equation}
\label{mu} \mu^3=\mu^2+(1-\epsilon)\mu+\epsilon.
\end{equation}
Differentiating this equation with respect to $\epsilon$ and setting
$\epsilon=0$, we find $d\mu/d\epsilon\big|_{\epsilon=0}$, and thus,
for small $\epsilon$ we have
$\mu(\epsilon)=\lambda-\frac{\lambda-1}{\lambda+2}\epsilon$. To
compare with the growth of the typical sequence (\ref{typical}), it
is useful to write $\langle x_n\rangle \sim e^{\gamma n}$ with
$\gamma=\ln \mu$. To first order in the disorder strength
$\epsilon$,
\begin{equation}
\label{gamma}
\gamma(\epsilon)=\gamma_0+\gamma_1\epsilon+\cdots
\end{equation}
with $\gamma_0=\beta_0$ and
$\gamma_1=\frac{1-\lambda}{\lambda(\lambda+2)}$.

To address the typical behavior, we introduce the ratio between two
successive elements in the sequence, \hbox{$r_n=x_n/x_{n-1}$}. The
random recursion rule (\ref{def}) implies that this ratio satisfies
the random map,
\begin{equation}
\label{map}
r_{n+1}=
\begin{cases}
   1+\frac{1}{r_n}                         & {\rm prob} \quad 1-\epsilon;\\
   1+\frac{1}{r_n}\cdot \frac{1}{r_{n-1}}  & {\rm prob} \quad
   \epsilon.
\end{cases}
\end{equation}

When there is no disorder, $\epsilon=0$, the ratio approaches the golden
number, $r_n\to \lambda$ as $n\to\infty$. Thus as the number of iterations of
the normal map (\ref{map}) grows indefinitely, the distribution of the ratio
approaches a delta function centered at the golden ratio, $P(r)\to
\delta(r-\lambda)$.

Generally, when $\epsilon>0$, the distribution $P(r)$ has a richer
structure, as shown below.  The Lyapunov exponent can be conveniently
expressed in terms of $P(r)$. Indeed, each sequence element is given
by the product
\begin{equation}
x_n=\prod_{j=2}^n r_j.
\end{equation}
With the exponential growth (\ref{typical}), the Lyapunov exponent
simply equals the expected value of the logarithm of the ratio,
\begin{equation}
\label{formula}
\beta=\langle\, \ln r\,\rangle=\int dr\, P(r)\, \ln r.
\end{equation}

At weak disorder, with a small probability $\epsilon$, an error
occurs. That is, the map $r_{n+1}=1+1/(r_nr_{n-1})$ is implemented.
As long as no errors occur, the ratio will essentially be equal to
$\lambda$. Then, when an error occurs, the ratio reduces to
$1+\lambda^{-2}$. Since the expected number of iterations before
another error occurs, $\epsilon^{-1}$, is very large, the ratio again
quickly approaches $\lambda$. This cycle continues ad-infinitum.

To characterize this process, we should understand how an error
evolves under the random map (\ref{map}). Thus, we consider the
following scenario: (1) Initially, the ratio equals the golden
number $\rho_0=\lambda$, (2) An error occurs at the very first
step, and (3) no further errors occur. Let $\rho_n$ be the value
of the ratio after $n$ iterations. Then $\rho_1=1+\lambda^{-2}$,
and using the relation $\lambda^2=\lambda+1$ we have
$\rho_1=\frac{2+\lambda}{1+\lambda}$. At further iterations, the
ratio follows the normal map $\rho_{n+1}=1+1/\rho_n$ and
therefore,
\begin{equation*}
\rho_1=\frac{2+\lambda}{1+\lambda}\,,\quad
\rho_2=\frac{3+2\lambda}{2+\tau}\,,\quad
\rho_3=\frac{5+3\lambda}{3+2\lambda}.
\end{equation*}
By induction, at the $n^{\rm th}$ iteration, the ratio can be expressed in
terms of the Fibonacci numbers
\begin{equation}
\label{rhon}
\rho_{n}=\frac{F_{n+2}+F_{n+1}\lambda}{F_{n+1}+F_n\lambda}.
\end{equation}
This series alternates around $\lambda$: $\rho_{2n+1}<\lambda$ while
$\rho_{2n}>\lambda$, but both the odd and the even sub-series quickly
converge to golden ratio, $\rho_n\to\lambda$ as $n\to\infty$.

This analysis characterizes how a single error affects the ratio. To
first order in the disorder strength $\epsilon$, the probability
that the value $\rho_n$ is observed equals
$\epsilon(1-\epsilon)^{n-1}$, reflecting the probability that one
error is made and then, no errors are made in the following $n-1$
iterations. Our first result is the distribution $P(r)$ for the
ratio to have the value $r$
\begin{equation}
\label{pr}
P(r)\to\epsilon\sum_{n=1}^\infty
(1-\epsilon)^{n-1}\delta\left(r-\rho_n\right),
\end{equation}
in the weak disorder limit $\epsilon\to 0$.

The calculation of the first order correction to the Lyapunov
exponent (\ref{pert}) is now straightforward. Substituting the
leading behavior in the weak disorder limit (\ref{pr}) into the
general formula (\ref{formula}) for the Lyapunov exponent we obtain
\begin{equation}
\lambda\to \epsilon\sum_{n=1} (1-\epsilon)^{n-1}\ln\rho_n.
\end{equation}
The sum is evaluated as follows
\begin{eqnarray*}
\lambda\to\epsilon\sum_{n=1}^\infty (1-\epsilon)^{n-1}\ln\lambda+
\epsilon\sum_{n=1}^\infty
(1-\epsilon)^{n-1}\ln\frac{\rho_n}{\lambda}.
\end{eqnarray*}
Performing the summation in the first term, we verify that
$\beta_0=\ln\lambda$. Keeping only the terms proportional to
$\epsilon$ in the second sum gives the leading correction in the
perturbation expansion of the Lyapunov exponent (\ref{pert}),
\begin{equation}
\label{correction} \beta_1=\sum_{n=1}^\infty
\ln\frac{\rho_n}{\lambda}.
\end{equation}
To perform this summation, we substitute the expression
(\ref{rhon}), and replace the upper limit with a large but finite
cutoff $N$, and then evaluate the $N\to\infty$ limit as follows
\begin{eqnarray*}
\beta_1&=&\lim_{N\to\infty} \ln
\frac{1}{\lambda^N}\frac{F_{N+2}+F_{N+1}\lambda}{1+\lambda} \\
&=&\ln \frac{2\lambda^2}{(\lambda+1)(\lambda+\lambda^{-1})}
\end{eqnarray*}
where in the second line we used Eq.~(\ref{fn}).  Using the equality
$\lambda^2=\lambda+1$, we arrive at our second main result
\begin{equation}
\label{beta1}
\beta_1=\ln \frac{2\lambda}{\lambda+2}\,.
\end{equation}
This correction is very close, but not identical, to that
corresponding to the average behavior (\ref{gamma}). As the difference
$\beta_1-\gamma_1\approx -0.00599897$ is negative, the typical growth
is slower than the average growth. This manifests the multiscaling
behavior that has been reported in other stochastic sequences
\cite{bk}. Generally, there is a multiscaling spectrum $\zeta_m$ that
characterizes the growth of the $m$th moments, $\langle
x_n^m\rangle^{1/m}\sim \exp(n\zeta_m)$. However, there is no obvious
relation between the Lyapunov exponent $\beta$ and the multiscaling
spectrum $\zeta_m$, e.g., $\beta\neq \gamma\equiv \zeta_1$.

\begin{figure}[t]
 \vspace*{0.cm}
 \includegraphics*[width=0.45\textwidth]{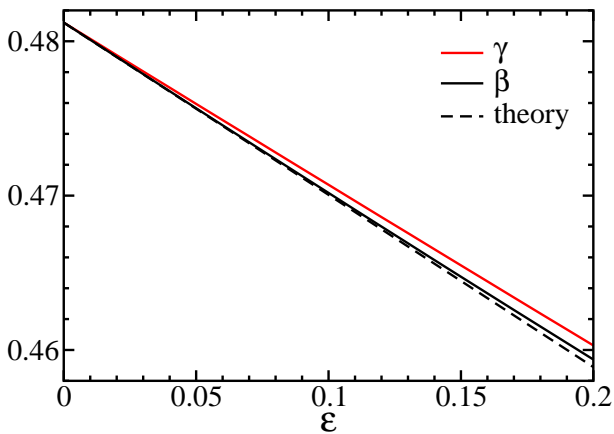}
 \caption{The weak disorder limit. Shown is the Lyapunov exponent
$\beta$, versus the disorder strength $\epsilon$.  Also shown for
reference are the parameter $\gamma$ characterizing the average
growth and the perturbation theory result (\ref{pert}) with
(\ref{beta1}).}
\end{figure}

We performed Monte Carlo simulations to verify these theoretical
predictions.  In the simulations, we followed the stochastic evolution
of the variable $r$.  This approach is advantageous for computation
because the ratios are bounded, in contrast with the explosive growth
in the sequence elements. The results presented here correspond to a
single Monte Carlo run with $10^9$ iterations.

There is a distinct but subtle difference between the typical and
the average growth as characterized by $\beta$ and $\gamma$,
respectively (Fig.~1). The two coincide in the limiting cases
$\epsilon=0$ and $\epsilon=1$ \cite{extremes}, and the discrepancy
is maximal, a mere $0.2\%$, at the midpoint $\epsilon=1/2$.

The numerical simulations show unambiguously that as the number of
iterations grows indefinitely, the ratio distribution approaches a
stationary distribution $P(r)$.   On Fig.~2, we display the
cumulative distribution \hbox{$G(r)=\int_{0}^r dr'\,P(r')$}.

The stationary distribution has a compact support, $r_{\rm min}<r<r_{\rm
  max}$. Indeed, the definition of the map (\ref{map}) implies the obvious
bounds $r_{\rm min}>1$ and $r_{\rm max}<2$.  The values $r_{\rm
  min}=(1+\sqrt{3})/2$ and $r_{\rm max}=\sqrt{3}$, consistent with the
numerical simulations results, are obtained from the following relations
\begin{subequations}
\begin{align}
\label{rel1}
r_{\rm max}&=1+\frac{1}{r_{\rm min}}\\
\label{rel2}
r_{\rm min}&=1+\frac{1}{1+r_{\rm max}}\,.
\end{align}
\end{subequations}
The first relation (\ref{rel1}) follows from the normal Fibonacci recurrence
$r_{n+1}=1+1/r_n$. The second relation (\ref{rel2}) follows from the altered
recurrence $r_{n+1}=1+1/(r_nr_{n-1})$ combined with ${\rm
  max}(r_nr_{n-1})=1+{\rm max}(r_{n-1})=1+r_{\rm max}$, that follows from the
normal recursion $r_{n+1}=1+1/r_{n-1}$.

\begin{figure}[t]
 \vspace*{0.cm}
 \includegraphics*[width=0.45\textwidth]{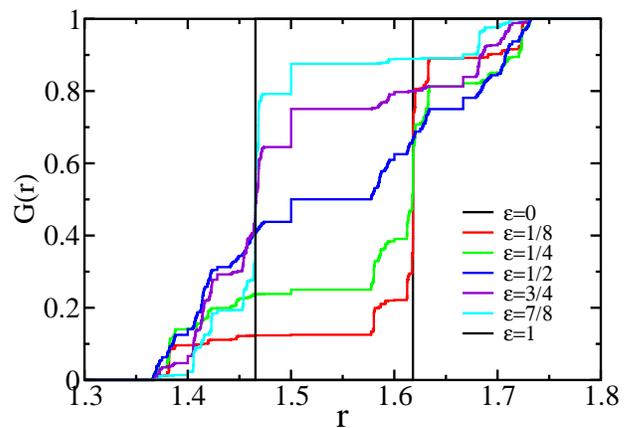}
 \caption{The cumulative distribution $G(r)$ versus the ratio $r$.}
\label{cumul}
\end{figure}

The distribution $P(r)$ consists of a set of delta functions, and
therefore, the cumulative distribution $G(r)$ has a devil's
staircase structure with infinitely many gaps. Generally, there is
a large gap in the interval $1/2<r<1+1/\sqrt{3}$. This gap arises
since the map (\ref{map}) transforms $(r_{\rm min},r_{\rm max})$
into the union of two subintervals, $(r_{\rm min},3/2)$ and
$(1+1/r_{\rm max},r_{\rm max})$. The bounding point $3/2$ is
obtained using reasoning similar to that used in the previous
paragraph. Restricting the map to the above subintervals one finds
that they are transformed into the union of four smaller
subintervals, etc. Hence the support of the invariant distribution
$P(r)$ is a Cantor-like fractal set and the cumulative
distribution therefore has a devil's staircase structure with an
uncountable number of singularities (Fig.~\ref{cumul}).

Thus far, we have addressed a specific modification of the Fibonacci
recurrence, namely, the one involving the index shift
$x_{n+1}=x_n+x_{n-2}$. But there are of course several other,
equally natural, modifications of the basic recursion rule. For
example, one may simply copy the last element $x_{n+1}=x_{n}$ or
alternatively, double it $x_{n+1}=x_{n}+x_{n}$. These two models are
analyzed along the same lines. For simplicity, we address the latter
case where the recursion relation is
\begin{equation}
\label{def2}
x_{n+1}=
\begin{cases}
   x_{n}+x_{n-1}   & {\rm prob} \quad 1-\epsilon,\\
   x_{n}+x_{n}   & {\rm prob} \quad \epsilon;
\end{cases}
\end{equation}
with the initial elements $x_1=x_2=1$. The corresponding random
map is
\begin{equation}
\label{map2}
r_{n+1}=
\begin{cases}
   1+\frac{1}{r_n}    & {\rm prob} \quad 1-\epsilon;\\
   2                  & {\rm prob} \quad \epsilon.
\end{cases}
\end{equation}
In contrast with (\ref{map}), when an error occurs, the ratio
\hbox{$r=2$} is independent of the previous element. Thus, error
events effectively reset the process anew. As a result, this
stochastic process is analytically tractable.

To characterize how an error propagates, we start with $\rho_1=2$
and use the recursion $\rho_{n+1}=1+1/\rho_n$ to obtain the first
few terms $\rho_2=3/2$, $\rho_3=5/3$, and $\rho_4=8/5$. In general,
\begin{equation}
\label{rhon2}
\rho_n=\frac{F_{n+2}}{F_{n+1}}.
\end{equation}
The ratio attains this value, $r=\rho_n$, when an error is followed
by $n-1$ normal recursion steps, and this occurs with probability
$\epsilon(1-\epsilon)^{n-1}$. Thus, the probability distribution of
the ratio is
\begin{equation}
\label{pr2}
P(r)=\epsilon\sum_{n=1}^\infty
(1-\epsilon)^{n-1}\delta\left(r-\rho_n\right)
\end{equation}
with $\rho_n$ given by (\ref{rhon2}). In contrast with the limiting
distribution (\ref{pr}), this result is now exact, because the
history prior to the most recent error event is irrelevant. The
distribution now has a countable set of singularities located at the
ratios $\rho_n$ of successive Fibonacci numbers. These singularities
 ``bunch'' near the golden ratio $\lambda$.

Substituting the probability distribution (\ref{pr2}) into the
Lyapunov formula (\ref{formula}) yields
\begin{equation}
\label{beta2}
\beta=\epsilon\sum_{n=1}^\infty
(1-\epsilon)^{n-1}\ln\frac{F_{n+2}}{F_{n+1}}.
\end{equation}
Again, the typical growth is slower than the average growth, as for
example, $\beta(1/2)\approx 0.571357$ while $\gamma(1/2)\approx
0.577049$ \cite{mu}.  The exact expression (\ref{beta2}) can be, in
principle, expanded as a power series in the disorder strength $\epsilon$,
viz. $\beta=\sum_{n\geq 0} \beta_n \epsilon^n$ with $\beta_n$ characterizing
the effect of $n$ errors. Of course, $\beta_0=\ln\lambda$. The lowest order
correction, which can be obtained either from Eq.~(\ref{beta2}) or from
Eq.~(\ref{correction}), is given by
\begin{equation}
\beta_1=\ln \frac{2\lambda+1}{\lambda+2}\,.
\end{equation}
One can also extract the next correction from Eq.~(\ref{beta2}); the result is
$\beta_2=\sum_{m\geq 0}
\ln\left[1+(-1)^m\lambda^{-2m-6}\right]$.

In summary, we introduced a class of random Fibonacci sequences
where with a fixed probability the classic rule is followed, but
otherwise, an alternate recursion occurs. We analyzed the weak
disorder limit and obtained the limiting distribution for the ratio
of consecutive sequence elements as well as the Lyapunov exponent.
We found that the typical growth is slower than the average growth.
We also showed that the cumulative distribution of the ratio of
consecutive elements has a devil's staircase structure.  An exact
solution for particularly simple alterations of the recursion rule
was obtained as well.

The above results raise a number of questions: Can the ratio
distribution and the Lyapunov exponent be obtained analytically in
general?  What are the locations of the singularities underlying the
distribution of the ratio? What is the probability that a given
integer belongs to the random Fibonacci sequence?

We have focused on the average and the typical sequence growth, but
further information is encoded in fluctuations with respect to the
typical behavior. Related studies on disordered systems suggest that
such fluctuations should obey Gaussian statistics \cite{bl} and our
preliminary numerical simulations support this. The corresponding
variance may be calculated using perturbation theory in the weak
disorder limit.

\smallskip
 We are indebted to Alan Newell for stimulating our interest in
imperfect sunflowers and thank Patrick Shipman for pointing out
the empirical data. We acknowledge US DOE grant W-7405-ENG-36 and
NSF grant CHE-0532969 for support of this work.


\begin{thebibliography}{99}
\bibitem{extremes}
     For $\epsilon=0,1$ the growth is of course deterministic and
     $\mu(0)=e^{\beta(0)}=\lambda$ and $\mu(1)=e^{\beta(1)}=1.465571232$.
\end{thebibliography}
\end{document}